\pdfoutput=0
\documentclass[11pt]{amsart}
\usepackage{amssymb,color}
\numberwithin{equation}{section}
\usepackage{amssymb}
\usepackage{graphicx}
\usepackage{url}

\theoremstyle{remark}
\newtheorem{rem}{Remark}[section]

\date{\today}

\begin{document}
\title[One-dimensional $q$-state modified Potts model]{One-dimensional $q$-state modified Potts model and its thermodynamic functions}
\author[Hasan Akin]{Hasan Ak\i n\\ Department of Mathematics, Faculty of Arts and Sciences,\\ Harran
University, Sanliurfa TR63050, Turkey,\\
e-mail: {\tt akinhasan@harran.edu.tr; akinhasan25@gmail.com}}
\date{\today }%
\begin{abstract}
Since its introduction, the Potts model has gained widespread
popularity across various fields due to its diverse applications.
Even minor advancements in this model continue to captivate
scientists worldwide, and small modifications often intrigue
researchers from different disciplines. This paper investigates a
one-dimensional \(q\)-state modified Potts model influenced by an
external magnetic field. By leveraging the transfer matrix method,
exact expressions are derived for key thermodynamic quantities,
including free energy, entropy, magnetization, susceptibility, and
specific heat capacity. Numerical analyses explore how these
thermodynamic functions vary with relevant parameters, offering
insights into the system's behavior. Additionally, the asymptotic
properties of these quantities are examined in the limiting cases
\(T \to 0\) and \(T \to \infty\). The findings contribute to a
deeper understanding of the model's thermodynamic characteristics
and highlight its potential applications across various
disciplines.
\\
\end{abstract}
\maketitle

\section{Introduction}\label{Intro}


Lattice spin systems encompass a wide range of models within
statistical mechanics \cite{Baxter}. While some directly
correspond to physical phenomena, others act as simplified analogs
of more intricate systems \cite{Preston}. The lattice structure
plays a fundamental role in studying spin systems. Since Potts
introduced the model \cite{Potts-1952}, systems involving three or
more spins on various grids have been extensively investigated
across numerous disciplines (statistical mechanics, mathematics,
material science, biology \cite{Graner-13}, quantum entanglement
and topological order \cite{Suzuki-1976}) under different names to
represent complex systems \cite{Bhat-2024}. The Potts model,
developed as a generalization of the Ising model \cite{Ising1925},
revolves around interacting spins, which in the Ising model are
limited to parallel or antiparallel states \cite{Potts-1952,Wu}.
In Ref. \cite{Cannas-1997}, the authors studied the
one-dimensional \( q \)-state Potts model (1D \( q \)-SPM) with
ferromagnetic pair interactions that decay with the distance \( r
\) according to a power-law function.

To investigate various fundamental properties of lattice systems
on specific grids, it is essential to analyze the model's
partition functions. Numerous methods exist for constructing
partition functions. In the 1D case, these can be expressed using
recurrence relations, but they are more generally derived using
the transition matrix method. This method was first introduced by
Kramers and Wannier in 1941 \cite{Kramers-I-1941,Kramers-II-1941}.
This approach enables the analysis of thermodynamic properties of
the lattice model via the partition function \cite{Cuesta-2004}.
In this study, we concentrate on determining the largest
eigenvalue of the transition matrix associated with a \(q\)-state
modified Potts model in the 1D setting (shortly, 1D $q$-SMPM).
Calculating the partition function for a chain of two-state
elements with nearest-neighbor interactions, known as the 1D Ising
model, is a standard exercise in many textbooks
\cite{Yurij-Entropy-2024}. This computation is typically carried
out using the transfer matrix method, which sums the exponential
terms derived from the Ising model Hamiltonian. In the same way,
the partition function for other 1D lattice models, such as the
Ising, Potts, or Hard-core models, is calculated using the same
method. The transfer matrix approach simplifies the summation of
exponential terms according to the specific Hamiltonian of each
model \cite{AMA-ZN-20,AMA-TMP-23,SS-TMP-23,WDC-19}. In Ref.
\cite{Creswick-1999}, Creswick and Kim utilized the microcanonical
transfer matrix to investigate the zeros of the partition function
of the \( Q \)-SPM. Potts models, studied under various names,
have been the focus of recent research. In Ref. \cite{Kry-2008},
Kryzhanovsky analyzed a modified version of the Potts model
featuring binarized synaptic coefficients.

By analyzing the eigenvalues of the transfer matrix, one can
directly obtain the partition function. The key distinction
between the combinatorial method (used by Ising) and the transfer
matrix method lies in how the Boltzmann weights are handled.
Recently, the transfer matrix method has yielded successful
results in studying the thermodynamic functions of 1D mixed-type
and mixed-spin lattice models \cite{A-Ax-2023,
Akin-2024,RR-2021,Akin-CJP-2024,Akin23Chaos,Akin-Chaos-2024}.

In \cite{Akin-Ulusoy-2023}, we examined various thermodynamic
properties of the \( q \)-SPM on the Cayley tree using the cavity
method. In Reference \cite{Akin-Ulusoy-2022}, the phase transition
problem was investigated by the cavity method on the Cayley tree
for the Potts model with finite number of spins. Similarly, in
\cite{Yang-CTP-2023}, Yang employed a modified recursion technique
to derive the exact partition function for a 1D Potts model with
free or periodic boundary conditions. This approach was then
compared to the transfer matrix method and the technique
introduced by Marchi and Vila. Furthermore, Chang and Shrock, in
\cite{Chang-23}, proposed measures of spin ordering for the \( q
\)-state ferromagnetic Potts model under a generalized external
magnetic field.

In this present paper, we introduce a 1D $q$-SMPM with an external
field. The article will be approached using two distinct methods.
First, the iterative equation system will be derived using the
cavity method, and the solutions of these equations identify the
Gibbs measure for the specified model. Second, the transition
matrix will be constructed based on the Markov chain rule, and the
largest eigenvalue of this matrix will be used to compute the
corresponding partition function. Using this partition function,
we will determine the model's free energy function. By taking
partial derivatives of the free energy function with respect to
the relevant parameters, we will derive exact formulas for
thermodynamic functions such as entropy, magnetization,
susceptibility, and specific heat capacity. Finally, numerical
calculations will be performed to plot the graphs of these
functions, allowing for comparisons with previous results.

The remainder of this study is organized as follows: Section
\ref{Sec:Prelimi-PPF} introduces the Hamiltonian that defines the
1D \(q\)-SMPM and constructs the transition matrix associated with
the linear equation system used to compute partial partition
functions for the 1D case. Sections \ref{SEC:Free-Energy} and
\ref{SEC:Thermo-Func} focus on deriving the model's basic
thermodynamic functions using the transition matrix for the 1D
\(q\)-SMPM. These functions are analyzed numerically, particularly
in relation to the specified parameters, and their peak points at
critical temperatures are explored. Finally, the concluding
section \ref{SEC:Conclusion} provides a comprehensive evaluation,
highlights key findings, and offers predictions to guide future
research.

\section{Construction of the partial partition functions and transition matrix}\label{Sec:Prelimi-PPF}

In this section, we introduce the essential concepts and results
that form the foundation of our analysis. To calculate the free
energy for the 1D $q$-SMPM with an external field, we use the
following Hamiltonian:
\begin{align}\label{HAM-1}
H(\sigma) &= -J \sum_{\langle x, y \rangle} \cos(\pi
\delta_{\sigma(x), \sigma(y)}) - h \sum_{\langle x, y \rangle}
\cos(\pi \delta_{\sigma(x), \sigma(y)}),
\end{align}
where \(\langle x, y \rangle\) denotes nearest-neighbor pairs, and
\(\delta_{\sigma(x), \sigma(y)}\) represents the Kronecker delta.

Let us go through a more specialized approach that considers the
discrete nature of the spin states. For spins \(\sigma(x),
\sigma(y) \in \Phi:=\{1, 2,\cdots,q\}\), the term \( \cos(\pi
\delta_{\sigma(x), \sigma(y)}) \) introduces different
interactions depending on whether the neighboring spins are equal
or not:
\begin{itemize}
    \item If \(\sigma(x) = \sigma(y)\), \(\delta_{\sigma(x), \sigma(y)} =
1\), so \(\cos(\pi \cdot 1) = -1\).
    \item If \(\sigma(x) \neq \sigma(y)\), \(\delta_{\sigma(x), \sigma(y)}
= 0\), so \(\cos(\pi \cdot 0) = 1\).
\end{itemize}
Thus, the interaction between neighbors \( x \) and \( y \) can be
summarized as:
\[
J \cos(\pi \delta_{\sigma(x), \sigma(y)}) =
\begin{cases}
-J, & \text{if } \sigma(x) = \sigma(y), \\
J, & \text{if } \sigma(x) \neq \sigma(y).
\end{cases}
\]

On 1D lattice $\mathbb{N}$, the interaction energy of a given spin
configuration \( \{\sigma_i\}\in \Phi^N \) is expressed as
\begin{align}\label{Interac-Ener}
E^{(N)}\left( \{ \sigma_i \} \right) &= -J \sum_{i=1}^{N} \cos(\pi
\delta_{\sigma_i , \sigma_{i+1} })-\frac{h}{\beta}
\sum_{i=1}^{N}\cos(\pi \delta_{\sigma_i, \sigma_{i+1}}).
\end{align}

As is typical, the partition function of the 1D-$q$SMPM is the sum
over all possible spin configurations:
\begin{align}\label{PF-1}
Z^{(N)}(\beta,h,J) = \sum_{\{\sigma_i\}\in  \Phi^N} \exp \left(
-\beta E^{(N)}\left(\{\sigma_i \} \right) \right),
\end{align}
where \(\beta=\frac{1}{k_B T}\) is the inverse temperature,
\(k_B\) is the Boltzmann constant, and \(T\) represents the
temperature.

Let \( Z_{\sigma(x)}^{(N)}(\beta,h,J) \) represent the partial
partition function of a subtree rooted at vertex \( x \), where
the spin at \( x \) is fixed to \( \sigma(x) \), expressed as a
function of the parameters \( \beta \), \( h \), and $J$. At the
boundary vertices (the last level of the tree), their contribution
to the partial partition function depends on the spin state, as
they have no neighboring vertices. For any initial vertex \( x \),
the value of \( Z_{\sigma(x)}^{(N)}(\beta,h,J) \) is calculated by
summing over all possible spin configurations of its consecutive
vertices. When \( \sigma(x) \in \Phi \), the contribution to \(
Z_{\sigma(x)}^{(N)}(\beta,h,J) \) is determined by whether the
spins of the consecutive vertices with or deviate from \(
\sigma(x) \). The recursion for \( Z_{\sigma(x)}^{(N)}(\beta,h,J)
\) is expressed through the following partial partition sums:
\begin{align}\label{PPF-1}
Z_{\sigma(x)}^{(N)}(\beta,h,J)&=\sum_{\sigma(y) \in \Phi} e^{\beta
J\cos(\pi \delta_{\sigma(x), \sigma(y)}) +h\cos(\pi
\delta_{\sigma(x), \sigma(y)})}Z_{\sigma(y)}^{(N-1)}(\beta,h,J),
\end{align}
where \( y\) represents the consecutive vertex of \( x \) on the
lattice $\mathbb{N}$.

The total partition function is given by:
\begin{align}\label{Tot-PF-2}
Z^{(N)}(\beta,h,J)
&=\sum_{\{states\}} e^{\beta J \sum_{i=1}^{N}\cos(\pi
\delta_{\sigma_i \sigma_{i+1}})+h\sum_{i=1}^{N}\cos(\pi
\delta_{\sigma_i \sigma_{i+1}})}\\\nonumber
&=\sum_{\sigma(x)\in\Phi }Z_{\sigma(x)}^{(n)}(\beta,h,J),
\end{align}
and the sum:
\[
\sum_{\{states\}} (\dots) = \prod_{i=1}^{N} \sum_{\sigma_i\in
\Phi} (\dots)
\]
indicates summation over the spin states at all sites. The total
partition function can be rewritten as a product of terms, each
depending on two neighboring spins, with periodic boundary
conditions \( \sigma_{N+1}=\sigma_1 \):
\[
Z^{(N)}(\beta,h,J)= \sum_{\{states\}} M(\sigma_1, \sigma_2)
M(\sigma_2, \sigma_3) \dots M(\sigma_{N-1}, \sigma_N) M(\sigma_N,
\sigma_1)
\]
where $M(\sigma, \sigma') = e^{(\beta J+h) \cos(\pi
\delta_{\sigma,\sigma'})}$ for $\sigma, \sigma'\in \Phi$.

From Equation \eqref{PPF-1}, recursive relations can be written
for \(Z_{\sigma(x)}^{(N)}(\beta,h,J)\), and it is convenient to
express them in matrix form as follows:
\[
\begin{pmatrix}
Z_{1}^{(N)} \\
Z_{2)}^{(N)}\\
\vdots\\
Z_{q}^{(N)}
\end{pmatrix}
=\mathbf{M}^{(q)}
\begin{pmatrix}
Z_{1}^{(N-1)} \\
Z_{2}^{(N-1)}\\
\vdots\\
Z_{q}^{(N-1)}
\end{pmatrix},
\]
where the transfer matrix \(\mathbf{M}^{(q)}\) is given by
\begin{align}\label{Matrix1}
\mathbf{M}^{(q)}&=\left(
\begin{array}{cccccc}
 e^{-h-J \beta } & e^{h+J \beta } & e^{h+J \beta } & \cdots & e^{h+J \beta } & e^{h+J \beta } \\
 e^{h+J \beta } & e^{-h-J \beta } & e^{h+J \beta } & \cdots & e^{h+J \beta } & e^{h+J \beta } \\
 e^{h+J \beta } & e^{h+J \beta } & e^{-h-J \beta } & \cdots & e^{h+J \beta } & e^{h+J \beta } \\
 \vdots &\vdots &\vdots &\ddots &\vdots &\vdots \\
 e^{h+J \beta } & e^{h+J \beta } & e^{h+J \beta } & \cdots & e^{-h-J \beta } & e^{h+J \beta
 }\\
 e^{h+J \beta } & e^{h+J \beta } & e^{h+J \beta } & \cdots & e^{h+J \beta } & e^{-h-J \beta }
\end{array}
\right).
\end{align}
%

In the upcoming sections, we will perform a numerical analysis of
these functions by computing the thermodynamic quantities
associated with the model, using the largest eigenvalue of the
matrix \(\mathbf{M}^{(q)}\).

\section{Free energy}\label{SEC:Free-Energy}
The free energy of a lattice model is essential for understanding
various thermodynamic and physical characteristics of the system
\cite{Suzuki-1976}. The free energy function can give many clues
about the stability, phase transition and equilibrium state of a
system \cite{Cuesta-2004}. In particular, the free energy governs
aspects such as thermodynamic stability, phase transitions,
entropy, specific heat capacity, and equilibrium properties. For
instance, first-order phase transitions are characterized by a
discontinuity in the first derivative of the free energy (such as
entropy or magnetization), while second-order phase transitions
correspond to discontinuities in the second derivative of the free
energy (like specific heat capacity or susceptibility)
\cite{Bish-2010}.

It is well-established that the partition functions associated
with lattice models can be computed using various approaches. On
the Bethe lattice, methods such as the cavity method
\cite{Akin-Ax-22, Akin23CJP} or the Kolmogorov consistency theorem
are applicable. For one-dimensional lattices, the transition
matrix method is the most commonly utilized technique. The
partition function can be expressed as a linear combination of the
\( N \)-th powers of the eigenvalues of the transfer matrix
\cite{Cuesta-2004}. Thus, in the thermodynamic limit, using the
total partition function \( Z^{(N)}(\beta,h,J) \) from
\eqref{Tot-PF-2}, we determine the free energy per site for the 1D
\( q \)-SMPM. As \( N \to \infty \), only the largest eigenvalue
dominates, resulting in the following expression:
\begin{align}\label{FE-FORMULA-1}
f(\beta,h,J)= -k_B T \lim_{N \to \infty} \frac{1}{N} \ln
Z^{(N)}(\beta,h,J),
\end{align}
where \( N \) is the number of vertices. For the sake of
simplicity, we will set the constant \( k_B \) equal to 1
throughout the paper.

By transfer matrix method, the total partition function
$Z^{(N)}(\beta,h,J)$ is determined by evaluating the trace of the
matrix product as follows:
\[
Z^{(N)}(\beta,h,J)=
\text{Tr}\left(\left(\mathbf{M}^{(q)}\right)^N\right) =
\sum_{j=1}^{q}\left(\lambda_{j}\right)^N,
\]
where \( \lambda_{j} \) are the eigenvalues of the transfer matrix
$\mathbf{M}^{(q)}$ given in \eqref{Matrix1} for an arbitrary spin
\( q\).

From Equation \eqref{Tot-PF-2}, this can be rewritten as:
\begin{align}\label{TPF-3a}
Z^{(N)}(\beta,h,J)= \sum_{j=1}^q \left(\lambda_{j}\right)^N =
\left(\lambda_{\max}\right)^N\left[ 1 + \sum_{j=1}^{q-1}
\left(\frac{\lambda_{j}}{\lambda_{\max}}\right)^N \right],
\end{align}
where \( \lambda_{j} \) are the eigenvalues of the transfer matrix
\( \mathbf{M}^{(q)} \), and  \( \lambda_{\max} =
\max\{\lambda_{1}, \lambda_{2}, \dots, \lambda_{q}\}\).

In the thermodynamic limit \( N \to \infty \), the partition
function simplifies to:
\begin{align}\label{TPF-3a1}
Z^{(N)}(\beta,h,J)= \left(\lambda_{\max}\right)^N,
\end{align}
as the contributions from smaller eigenvalues become negligible.
It is well established that the critical behavior of the model is
governed by this largest eigenvalue in the thermodynamic limit as
\( N \to \infty \) \cite{AMA-TMP-23,WDC-19,A-Ax-2023,Akin-2024}.
From Equation \eqref{TPF-3a1}, the bulk free energy is governed by
the largest eigenvalue of the transfer matrix, and is given by:
\[
f(\beta,h,J)= -\frac{1}{\beta}\lim_{N \to \infty} \frac{1}{N} \ln
Z^{(N)}(\beta,h,J)= -\frac{1}{\beta} \ln \lambda_{\max}.
\]
The eigenvalues of $\mathbf{M}^{(q)}$ are derived based on the
structure of the matrix, taking into account its symmetry and the
constant diagonal values. By means of the characteristic equation
det$\left(\mathbf{M}^{(q)}-\lambda I\right)=0$, the eigenvalues of
the symmetric matrix \( \mathbf{M}^{(q)} \) are obtained as
\begin{align}\label{TMEVs1}
\lambda_i =
\begin{cases}
e^{-h - J\beta} \left(1 - e^{2h + 2J\beta}\right), & \text{for } 1 \leq i \leq q-1, \\
e^{-h - J\beta} \left(1 + (q-1)e^{2h + 2J\beta}\right), &
\text{for } i = q.
\end{cases}
\end{align}

From Equation \eqref{TMEVs1}, it is evident that \(\lim_{N \to
\infty}\left(\lambda_i / \lambda_q\right)^N =0\) since \(
\lambda_i / \lambda_q < 1 \) for $1\leq i\leq q-1$. Consequently,
considering the largest eigenvalue \( \lambda_q = \lambda_{\max} =
e^{-h - J\beta} \left(1 + (q-1)e^{2h + 2J\beta}\right) \) as given
in \eqref{TMEVs1}, the free energy function for the 1D-\( q \)SPM,
expressed in terms of \( J \), \( h \), \( \beta \), and \( q \),
is determined as:
\begin{align}\label{FEF-q1}
f(\beta,h,J,q)&= -\frac{1}{\beta}\ln \left( \frac{1 + (q-1)e^{2h +
2J\beta}}{e^{h+J\beta}} \right).
\end{align}

This function is also referred to as the Gibbs free energy per
site \cite{Chang-23}. For the disordered phase, we expect each
spin configuration to contribute equally, leading to a higher
entropy and potentially higher free energy, depending on  \( \beta
\), \( h \), \( J \) and $q$, respectively. This recursive
approach offers a way to approximate \( f \) numerically and
analyze phase behavior on the model with the given Hamiltonian
\eqref{HAM-1} and spin set \(\Phi\).

\begin{figure}[!htbp]
\centering
\includegraphics[width=140mm]{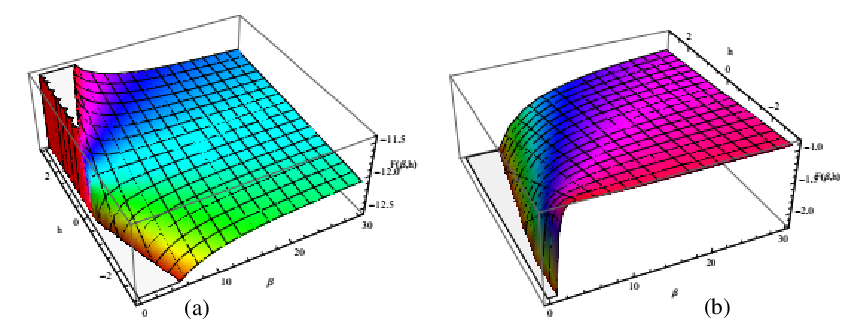}
\caption{(Color online) 3D plots of the free energy function
\eqref{FEF-q1} for the range \(\beta \in [0.001, 30]\) and \(h \in
[-3, 3]\): (a) for \(J = -12, q = 16\); (b) for \(J = 0.95, q =
16\).}\label{Graph-3DFE}
\end{figure}

Figures \ref{Graph-3DFE} display 3D representations of the free
energy function within the ranges \(\beta \in [0.001, 30]\) and
\(h \in [-3, 3]\). These graphs illustrate that the key
determinant is the \(J\) parameter. For antiferromagnetic cases
(\(J < 0\)), peaks are visible on the surface Figure
\ref{Graph-3DFE}(a), whereas for ferromagnetic cases (\(J
> 0\)), the surface gradually increases with \(\beta\) and
exhibits a smooth appearance Figure \ref{Graph-3DFE}(b).

\begin{figure}[!htbp]
\centering
\includegraphics[width=75mm]{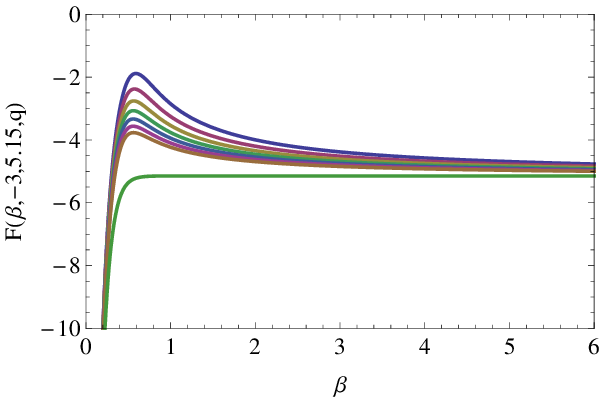}\ \ \ \
\includegraphics[width=75mm]{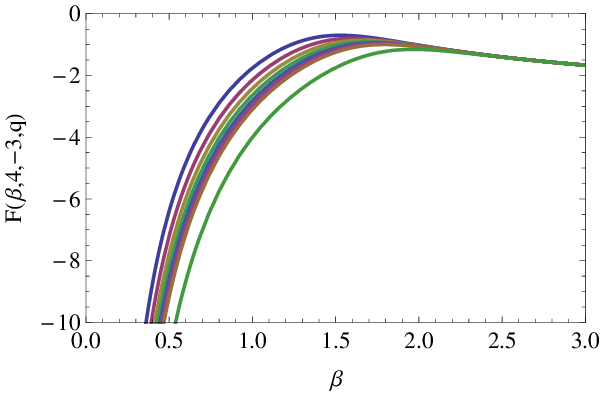}
\caption{(Color online) Graphs of free energy given in
\eqref{FEF-q1} (left) for $h=-3,J=5.15$ (right) for $h=4,J=-3$.
}\label{Graph-FE}
\end{figure}

In Fig. \ref{Graph-FE}, the free energy functions associated with
the 1D $q$-SMPM are plotted as a function of \(\beta\) for the
model with \(q = 3, 4, 5, 6, 7, 8, 9, 21\). The graphs are
generated for selected values of the coupling constants \(J\) and
\(h\). It is observed that the free energy function decreases with
increasing \(q\). Specifically, within certain intervals of
\(\beta\), the following inequality holds: $f(\beta, h, J, 21) <
f(\beta, h, J, 9) < f(\beta, h, J, 8) < f(\beta, h, J, 7) <
f(\beta, h, J, 6) < f(\beta, h, J, 5) < f(\beta, h, J, 4) <
f(\beta, h, J, 3).$ In \cite{Akin-Ulusoy-2023}, the authors
studied the classical \(q \)-SPM on the Cayley tree, considering
competing NN interactions and prolonged NNN interactions. They
also explored the behavior of various thermodynamic functions
associated with the model. In the current work, we observe that
the graphs presented in Figure \ref{Graph-FE} for the 1D $q$-SMPM
display similar patterns to those reported in
\cite{Akin-Ulusoy-2023}. Notably, for \( J = 5.15, h = -3 \), the
free energy functions show peaks in specific regions, whereas for
\( J = -3, h = 4 \), they exhibit a smooth and continuous
behavior. Note that the figures were drawn with the help of
Mathematica \cite{Mathematica}.

\begin{rem}
It is worth mentioning that the free energy function for the model
can also be determined using the total partition function derived
in equation \eqref{Tot-PF-2}.
\end{rem}

\section{Thermodynamic Functions}\label{SEC:Thermo-Func}

In this section, we extend our analysis of the 1D $q$-SMPM by
deriving additional thermodynamic properties, such as entropy,
magnetization, susceptibility, and specific heat capacity, using
the free energy function defined in \eqref{FEF-q1}. Furthermore,
we investigate how these properties depend on the parameters
\(\beta\), \(h\), \(J\), and \(q\), and we perform a numerical
analysis to study the behavior of their corresponding graphs.

\subsection{The entropy}
It is important to note that the derivatives of the free energy \(
F(\beta, h, J, q) \) with respect to different parameters provide
insights into the thermodynamic variables. Specifically, the
derivative of the free energy function \( f(\beta, h, J, q) \)
with respect to temperature \( T \) is related to the system's
entropy, as outlined by Georgii \cite{Georgii} (also see
\cite{Akin-Ulusoy-2023,Akin-Ulusoy-2022} for details). So, the
entropy of the 1D $q$-SMPM is given by:
\begin{align}\label{Entropy1}
S(\beta,h,J,q)&= - \frac{\partial F(\beta,h,J,q)}{\partial T} =
\beta^2 \frac{\partial F(\beta,h,J,q)}{\partial \beta}.
\end{align}

From the expressions \eqref{FEF-q1} and \eqref{Entropy1}, the
entropy \(S(\beta,h, q,J )\) is obtained as:
\begin{align}\label{Entropy1a}
S(\beta,h,J,q)&= \frac{J \left(1 -(q-1)e^{2(h + J \beta)}\right)
\beta + \left(1 + (q-1)e^{2(h + J \beta)}\right) \ln\left[e^{-h -
J \beta} \left(1 + (q-1)e^{2(h + J \beta)}\right)\right]}{1 +
(q-1)e^{2(h + J \beta)}}\\\nonumber &=\frac{J \left(1-e^{2
\left(h+\frac{J}{T}\right)} (q-1))\right)T^{-1}+\left(1+e^{2
\left(h+\frac{J}{T}\right)} (q-1)\right) \ln\left[e^{-\frac{J+h
T}{T}} \left(1+e^{2 \left(h+\frac{J}{T}\right)}
(q-1)\right)\right]}{1+e^{2 \left(h+\frac{J}{T}\right)} (q-1))}.
\end{align}

\begin{figure}[!htbp]
\centering
\includegraphics[width=140mm]{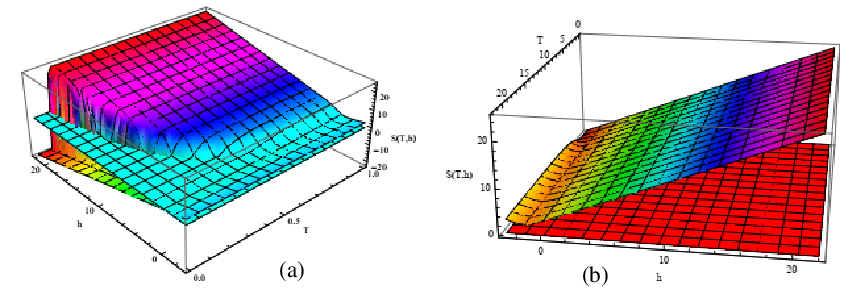}
\caption{(Color online) Three-dimensional (3D) plots of the
entropy function \eqref{Entropy1a} within the range \(T\in
[0.001,2]\) and \(h \in [-3,18]\): (a) for $J = -2, q = 17$ (b)
for $J =8, q =22$ within the range \(T\in [0.001,22]\) and \(h \in
[-3,22]\)}\label{Graph-3DENTROPY}
\end{figure}

The formula \eqref{Entropy1a} expresses the entropy in terms of
the parameters \(h\) and \(\beta\), incorporating the behavior of
the system based on these variables. So, one examines the effects
of \(J\) (interaction strength) and $q$ (number of states) on the
system's thermodynamic behavior.

In figures \ref{Graph-3DENTROPY}, we have plotted two 3D graphs
based on the formula in \eqref{Entropy1a}, with \( \beta = 1/T \)
as a function of \( h \) and \( T \). From these plots, we
observed that \( J \) has a greater influence than the \( q
\)-spin value. In the antiferromagnetic region (\( J < 0 \))
\ref{Graph-3DENTROPY}(a), the entropy function shows significant
variation, with some areas exhibiting negative values. In
contrast, in the ferromagnetic region (\( J > 0 \))
\ref{Graph-3DENTROPY}(b), the entropy function appears more
plane-like. In these graphs, we have also included the 0 plane in
the three-dimensional coordinate system to illustrate the regions
where the entropy is positive. This allows us to identify the
areas where the entropy value is greater than zero.

\begin{figure}[!htbp]
\centering
\includegraphics[width=70mm]{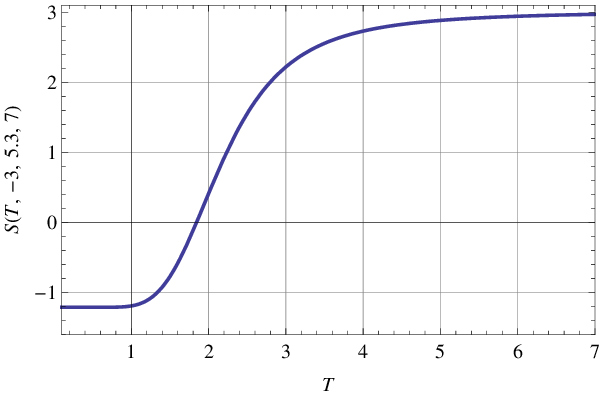}\ \ \ \
\includegraphics[width=70mm]{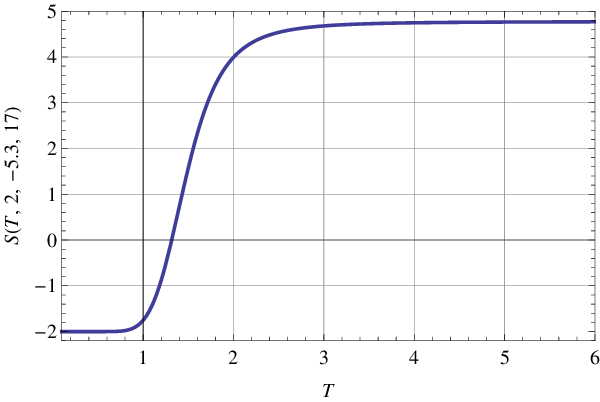}
\caption{(Color online) Entropy function plots \eqref{Entropy1a}
as a function of temperature \(T\): (left) for \( h = -3 \), \( J
= 5.3 \), \( q = 7 \); (right) for \( h = 2 \), \( J = -5.3 \), \(
q = 17 \), 22.}\label{Graph-2D-ENTROPY}
\end{figure}

Figure \ref{Graph-2D-ENTROPY} shows the graphs of the entropy
function as a function of temperature \( T \) for specific values
of \( h \), \( J \), and \( q \). A notable observation is that
the entropy value is negative at low temperatures.

\subsection{The magnetization per spin}
The magnetization \( m\left(\beta,h,J,q\right)\) is derived within
the framework of the canonical ensemble using equation
\eqref{TPF-3a}. To determine the magnetization for arbitrary
values of the external magnetic field \( h\), we will utilize
Onsager's exact solution for zero-field magnetization as a
reference. Following the methodology commonly used in the
literature \cite{AMA-ZN-20,AMA-TMP-23,A-Ax-2023}, the
magnetization per spin is determined by taking the partial
derivative of the free energy function with respect to the
external magnetic field \( h \).

Using the largest eigenvalue (\(\lambda_{\text{max}}\)) of the
transition matrix $\mathbf{M}^{(q)}$, we can compute the
magnetization per spin as
\begin{align}\label{Mag-Formula1}
m\left(\beta,h,J,q\right)&=- \frac{\partial f\left(\beta, h,
q,J\right)}{\partial h}=\frac{-1+e^{2 (h+J \beta )}
(q-1)}{\left(1+e^{2 (h+J \beta )} (q-1)\right) \beta }.
\end{align}

\begin{figure}[!htbp]
\centering
\includegraphics[width=140mm]{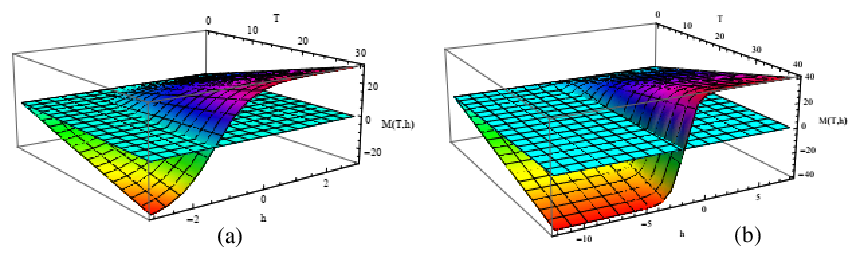}
\caption{(Color online) Three-dimensional plots of the
magnetization function \eqref{Mag-Formula1} are presented for the
following parameter ranges: (a) \( J = -2, q = 17 \) with \( T \in
[0.001, 30] \) and \( h \in [-3, 3] \); (b) \( J = 8, q = 22 \)
with \( T \in [0.001, 40] \) and \( h \in [-8, 8]\).
}\label{Graph-3DMAGNET}
\end{figure}

\begin{figure}[!htbp]
\centering
\includegraphics[width=80mm]{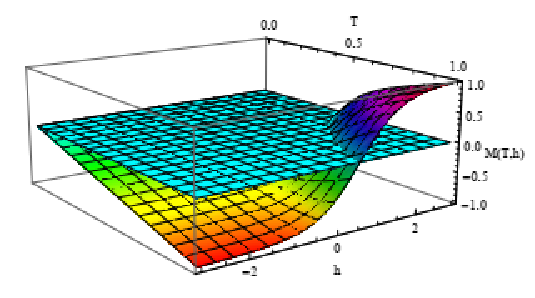}\ \ \ \
\caption{(Color online) Three-dimensional plot of the
magnetization function \eqref{Mag-Formula1} is presented for the
following parameter ranges: \( J = -2, q = 17 \) with \( T \in
[0.001,1] \) and \( h \in [-3, 3] \). }\label{Graph-3DMAGNET-T-1}
\end{figure}

In Equation \eqref{Mag-Formula1}, the free energy attains its
extreme value with respect to \( h \) when \( h = -\frac{1}{2}\ln
\left(e^{2\beta J}(q-1)\right) \). In Fig. \ref{Graph-3DMAGNET},
we present three-dimensional plots of the magnetization function
\eqref{Mag-Formula1} for the following parameter ranges: (a) \( J
= -2, q = 17 \) with \( T \in [0.001, 30] \) and \( h \in [-3, 3]
\); (b) \( J = 8, q = 22 \) with \( T \in [0.001, 40] \) and \( h
\in [-8, 8]\). From the graphs, it can be observed that the
magnetization function increases as \( h \) increases, exhibiting
both negative and positive values. Additionally, as the
temperature approaches zero, the magnetization becomes entirely
negative (see Fig. \ref{Graph-3DMAGNET-T-1}).

\subsection{Susceptibility}
The susceptibility of lattice models has been the subject of
extensive research, with a particular focus on its relationship
with other thermodynamic functions, especially the Curie
temperature. In Ref. \cite{MPR-21}, Melnikov, Paradezhenko, and
Reser applied dynamic spin fluctuation theory to calculate the
magnetic susceptibility of ferromagnetic metals above the Curie
temperature \( T_c \), and they made a thorough comparison with
experimental data. The susceptibility is often essential for
understanding the magnetic properties of materials \cite{MPR-21}.
By taking the first derivative of the free energy, as presented in
\eqref{FEF-q1}, calculated using the partition function for the
lattice model, with respect to the external magnetic field \( h
\), the magnetization per site is obtained. The partial derivative
of the function in \eqref{FEF-q1} with respect to \( h \) defines
the model's susceptibility \cite{Chang-23}. This function helps
identify the critical temperatures \( T_c \) and determine the
regions where phase transitions occur in the system
\cite{AMA-ZN-20,AMA-TMP-23}. Typically, sharp peaks marking the
transition from a disordered to an ordered state provide valuable
insight into the presence of a phase transition.

Magnetic susceptibility \(\chi\left(\beta,h,J,q\right)\) measures
the response of magnetization \( m\left(\beta, h, q,J\right)\) to
an external magnetic field \( h \), and is given by:
\begin{align*}\label{Susc1a}
\chi\left(\beta,h,J,q\right)= \left( \frac{\partial m\left(\beta,
h, q,J\right)}{\partial h} \right)_T
\end{align*}

In systems like  the Curie-Weiss model, susceptibility near the
Curie temperature \( T_c \) follows:
\[
\chi(T) = \frac{C}{T - T_c}
\]
where \( C \) is the Curie constant.

%

Consequently, by taking the second partial derivative of the free
energy function with respect to the field variable \( h \), we
arrive at the following expression:
\begin{align}\label{Susceptibility}
\chi\left( h,\beta,J, q\right)&= - \frac{\partial^2 f\left(
h,\beta,J, q\right)}{\partial h^2} = \frac{4 e^{2 (h+J \beta)}
(-1+q)}{\left(1+e^{2 (h+J \beta)} (-1+q)\right)^2 \beta}.
\end{align}

\begin{figure}[!htbp]
\centering
\includegraphics[width=70mm]{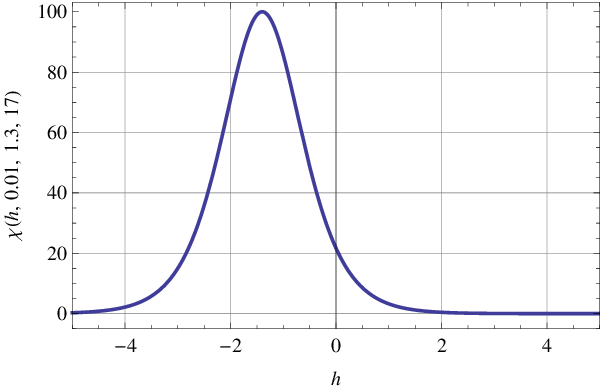}\ \ \ \
\includegraphics[width=70mm]{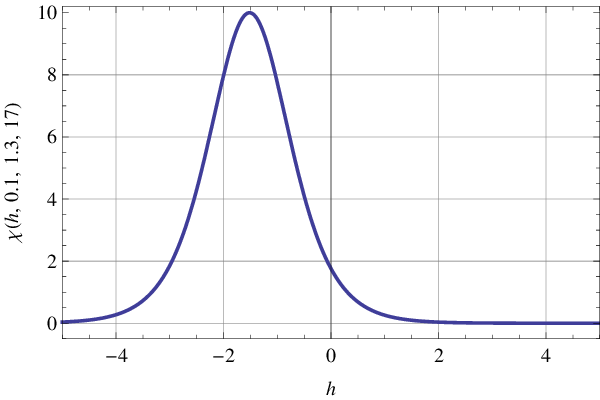}\ \
\includegraphics[width=70mm]{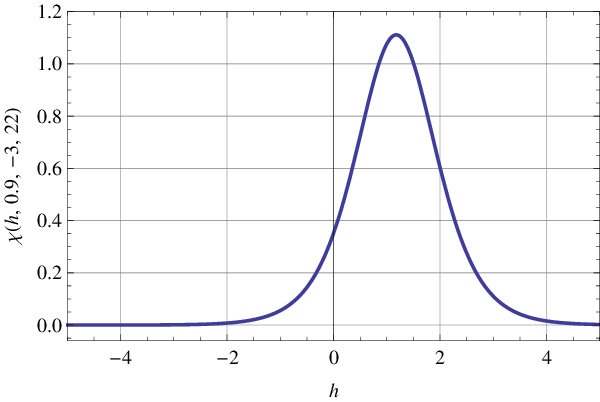}\ \ \ \
\includegraphics[width=70mm]{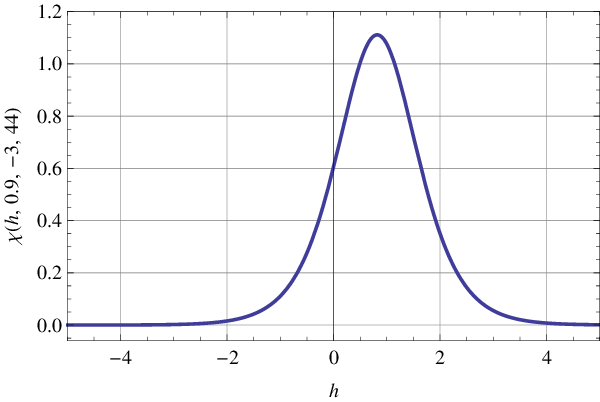}
\caption{(Color online) Graphs of the magnetic susceptibility \(
\chi(h, \beta, J, q) \) from equation \eqref{Susceptibility} are
shown as a function of the magnetic field \( h \) for the
following parameter sets: \( \beta = 0.01, J = 1.3, q = 17 \); \(
\beta = 0.1, J = 1.3, q = 17 \); \( \beta = 0.9, J = -3, q = 22
\); and \( \beta = 0.9, J = -3, q = 44 \),
respectively.}\label{Graph-SUS}
\end{figure}

In Figure \ref{Graph-SUS}, the magnetic susceptibility from
equation \eqref{Susceptibility} is plotted as a function of the
magnetic field \( h \) for the following parameter sets: \( \beta
= 0.01, J = 1.3, q = 17 \); \( \beta = 0.1, J = 1.3, q = 17 \); \(
\beta = 0.9, J = -3, q = 22 \); and \( \beta = 0.9, J = -3, q = 44
\). All the graphs display peaks. As the temperature increases (\(
\beta = 0.01; T=100 \)), the magnetic susceptibility also
increases. Notably, in the antiferromagnetic case with \( J = -3
\), the peak shifts to the right. Another key observation is the
significant influence of the \( q \)-spin value on the graph
behavior. Comparing the graphs for \( q = 22 \) and \( q = 44 \),
while keeping \( \beta = 0.9 \) and \( J = -3 \) constant, reveals
substantial differences between the two cases. In
\cite{AMA-ZN-20}, the authors demonstrated that the peaks of
susceptibility indicate a higher maximum of \(\chi(T, H)\) at
lower temperatures. Contrary to their findings, our model reveals
that the peak value of magnetic susceptibility increases with
rising temperature.

In summary, for the 1D $q$-SMPM, the surface curves corresponding
to thermodynamic functions along the $T$-axis (or $\beta$) and
$h$-axis provide a powerful way to study the intricate behaviors
of systems in statistical mechanics and thermodynamics, including
phase transitions, critical points, and the response to external
fields.
%
\subsection{Specific heat capacity}
Another important critical exponent of the 1D \( q \)-SMPM is
associated with the heat capacity, defined as:
\[
C(h, \beta, J, q) = k_B \beta^2 \frac{\partial^2}{\partial
\beta^2} \ln \lambda_{\text{max}}.
\]
Alternatively, it can be expressed as:
\begin{align}\label{capacity}
C(T,h,J, q)&= -T \frac{\partial^2 F}{\partial T^2}=\frac{4 e^{2
h+\frac{2 J}{T}} J^2 (q-1)}{\left(1+(q-1)e^{2
\left(h+\frac{J}{T}\right)}\right)^2 T^2}.
\end{align}

\begin{figure}[!htbp]
\centering
\includegraphics[width=155mm]{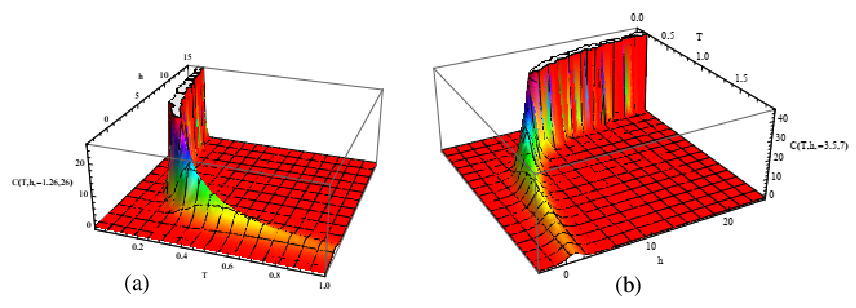}
\caption{(Color online) Three-dimensional plots of the heat
capacity \eqref{Mag-Formula1} are presented for the following
parameter ranges: (a) \( J = -1.26, q = 26 \) with \( T \in
[0.001,1] \) and \( h \in [-8, 15] \); (b) \( J = -3.5, q = 7 \)
with \( T \in [0.001, 1.9] \) and \( h \in [-3,25]\).
}\label{Graph-Capacity}
\end{figure}

In Figure \ref{Graph-Capacity}, we present three-dimensional plots
of the heat capacity \eqref{Mag-Formula1} for the following
parameter ranges: \( J = -1.26, q = 26 \) with \( T \in [0.001, 1]
\) (\ref{Graph-Capacity}(a)) and \( h \in [-8, 15] \), and \( J =
-3.5, q = 7 \) with \( T \in [0.001, 1.9] \) and \( h \in [-3, 25]
\) (\ref{Graph-Capacity}(b)). These plots clearly demonstrate that
at critical temperatures \( T \), the function either attains a
local maximum or displays asymptotic behavior.



\begin{figure}[!htbp]
\centering
\includegraphics[width=65mm]{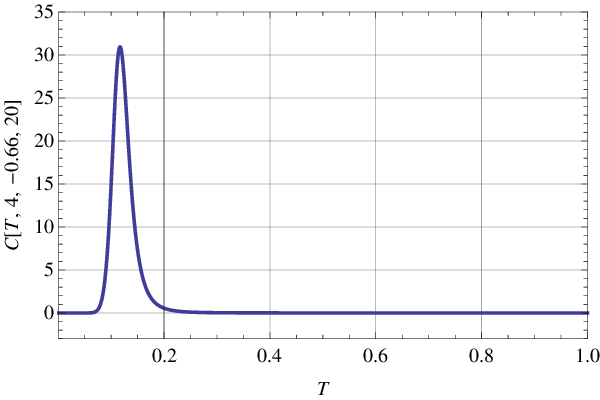}\ \ \ \
\includegraphics[width=65mm]{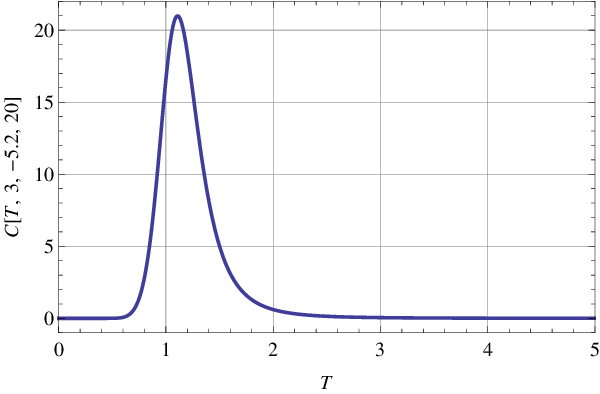}
\includegraphics[width=65mm]{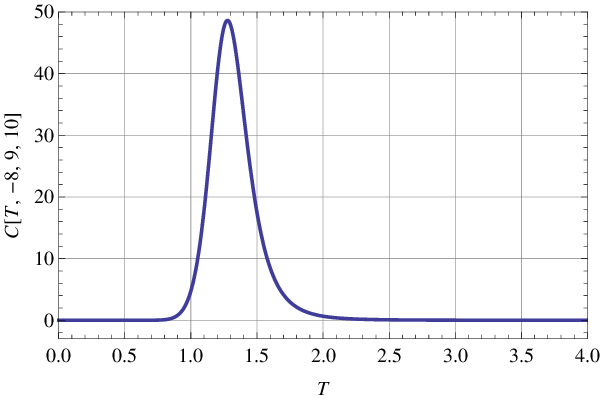}
\caption{(Color online) Temperature dependence of the heat
capacity function \eqref{capacity}: (a) for \(h=4, J=-0.66,
q=20\); (b) for \(h=3, J=-5.2, q=20\); and (c) for \(h=-8, J=9,
q=10\).}\label{Graph-2DCapacity}
\end{figure}

In Figure \ref{Graph-2DCapacity}, we display behavior of the heat
capacity function plots \eqref{capacity} as a function of
temperature \(T\): (left) for \(h=4, J=-0.66, q=20\); (right) for
\(h=3, J=-5.2, q=20\). The locations of the peaks in the function
change depending on the values of \(h\), \(J\), and \(q\). As
shown in the figures, the parameters influence the heat capacity
behavior. In the first two cases, where the system is considered
antiferromagnetic (\(J < 0\)), the shapes can be adjusted based on
the positive and negative external magnetic field \(h\).
Additionally, as \(q\) decreases, the value of the peak increases.


\section{Conclusions}\label{SEC:Conclusion}
As demonstrated in \cite{Akin-Ulusoy-2023}, constructing partial
partition functions for the \(q\)-state Potts model on the Cayley
tree results in non-linear recurrence equation systems, making it
impractical to directly derive the transition matrix from these
equations. This limitation necessitates the exploration of
alternative methods beyond the cavity method. In
\cite{Akin-Ax-22}, the free energy for an interactive Ising model
with nearest and extended next-nearest neighbor interactions on a
Cayley tree was successfully calculated using the cavity method.

In this study, we introduced a 1D \(q\)-SMPM under the influence
of an external field. The analysis was performed using two
complementary approaches. First, we employed the cavity method to
derive a system of iterative equations, whose solutions determine
the Gibbs measure for the model. In the one-dimensional case, this
system is linear, allowing us to directly obtain the corresponding
transition matrix. Second, we constructed the transition matrix
using the Markov chain rule and utilized its largest eigenvalue to
calculate the partition function. This partition function was then
used to derive the model's free energy function.

By taking partial derivatives of the free energy function with
respect to relevant parameters, we obtained exact expressions for
key thermodynamic quantities, including free energy, entropy,
magnetization, susceptibility, and specific heat capacity.
Additionally, numerical calculations were performed to visualize
these thermodynamic functions and compare them with previously
reported results.


It is well-known that one-dimensional models with short-range
interactions do not exhibit phase transitions when the single-spin
space is finite, as formalized in van Hove's theorem
\cite{Cuesta-2004}. However, Khakimov proved in
\cite{Khakimov-JSTAT-23} that a phase transition can occur in the
one-dimensional  SOS model with countable state under the
influence of an external field. In our work, the existence of a
phase transition for the one-dimensional \(q\)-SMPM in the
countable case will be examined in future research. Accordingly,
our future studies will investigate the phase transition behavior
of this model on finite-order Cayley trees using the cavity method
and the Kolmogorov consistency theorem. Additionally, the physical
relevance of this model remains an open question that warrants
further investigation.

The application of the transition matrix method to analyze
thermodynamic functions in lattice models is a fundamental concept
commonly introduced in undergraduate statistical mechanics
courses. We believe the findings of this study will be of
significant interest to researchers in the field of statistical
mechanics.

\begin{flushleft}
\textbf{Acknowledgment}\\
Part of this research was conducted during my time at ICTP in 2023, and I am grateful for the support I received from ICTP.
\end{flushleft}

\begin{flushleft}
\textbf{Funding}\\ This research received no external funding.
\end{flushleft}

\begin{flushleft}
\textbf{Data Availability Statement}\\ The statistical data
presented in the article do not require copyright. They are freely
available and are listed at the reference address in the
bibliography.
\end{flushleft}

\begin{flushleft}
\textbf{Conflicts of Interest}\\ The author declare no conflict of
interest.\end{flushleft}

\end{document}